\documentclass[a4paper]{article}

\usepackage{INTERSPEECH2021}
\usepackage{hyperref}

\title{Voice Activity Projection: Self-supervised Learning of Turn-taking Events}
\name{Erik Ekstedt, Gabriel Skantze}
\address{KTH Speech, Music and Hearing \\
  Stockholm, Sweden
}
\email{erikekst@kth.se, skantze@kth.se}

\begin{document}

\maketitle
\begin{abstract}
The modeling of turn-taking in dialog can be viewed as the modeling of the dynamics of voice activity of the interlocutors. We extend prior work and define the predictive task of Voice Activity Projection, a general, self-supervised objective, as a way to train turn-taking models without the need of labeled data. We highlight a theoretical weakness with prior approaches, arguing for the need of modeling the dependency of voice activity events in the projection window. We propose four zero-shot tasks, related to the prediction of upcoming turn-shifts and backchannels, and show that the proposed model outperforms prior work. 
\end{abstract}


\section{Introduction}
Turn-taking is the fundamental ability of humans to organize spoken interaction, i.e., to coordinate who the current speaker is, in order to avoid the need for interlocutors to listen and speak at the same time \cite{sacks:74}. A dialog can be viewed as a sequence of turns, constructed through the joint activity of turn-taking between the two speakers. A turn refers to segments of activity where a single speaker controls the direction of the dialog. In conversational systems, turn-taking has traditionally been modeled using threshold policies which recognize silences longer than a chosen duration as transition-relevant places. Although these types of models are commonly used, it is well known that they are insufficient for modeling human-like turn-taking \cite{Skantze2021}, since silences inside of a speaker turn are often longer than silences between turns. Thus, such policies will either result in interruptions or sluggish responses.

However, since pauses, overlaps and interruptions are very common phenomena, and since turn-taking is often optional, it is hard come up with objective criteria for what a turn really is, and how to automatically identify turns in dialog data \cite{Skantze2021}. One approach to make data-driven modeling of turn-taking feasible, is to instead model the dynamics of the voice-activity (VA) of the dialog participants (i.e., the binary notion of whether a person is speaking or not). From the VA, we can then identify certain events related to turn-taking. When VA from one speaker transitions to VA from the other speaker, we can identify a SHIFT, and when there is a silence within the VA of a single speaker (without VA from the other speaker), there is a HOLD. During a turn controlled by one speaker there may also be shorter, isolated VA from the other speaker, which roughly corresponds to the phenomenon called \textit{backchannels} \cite{Yngve}, i.e., shorter feedback vocalizations such as ``mhm". These turn-taking events are very important to identify for a conversational system. For example, it should be able to tell whether a mutual silence should be identified as a SHIFT or a HOLD. It is also important to identify when it is appropriate to backchannel, and whether the start of a VA is likely to constitute a backchannel or a longer contribution (i.e., a proper turn shift). 
\begin{figure}[t]
  \centering
  \includegraphics[width=\linewidth]{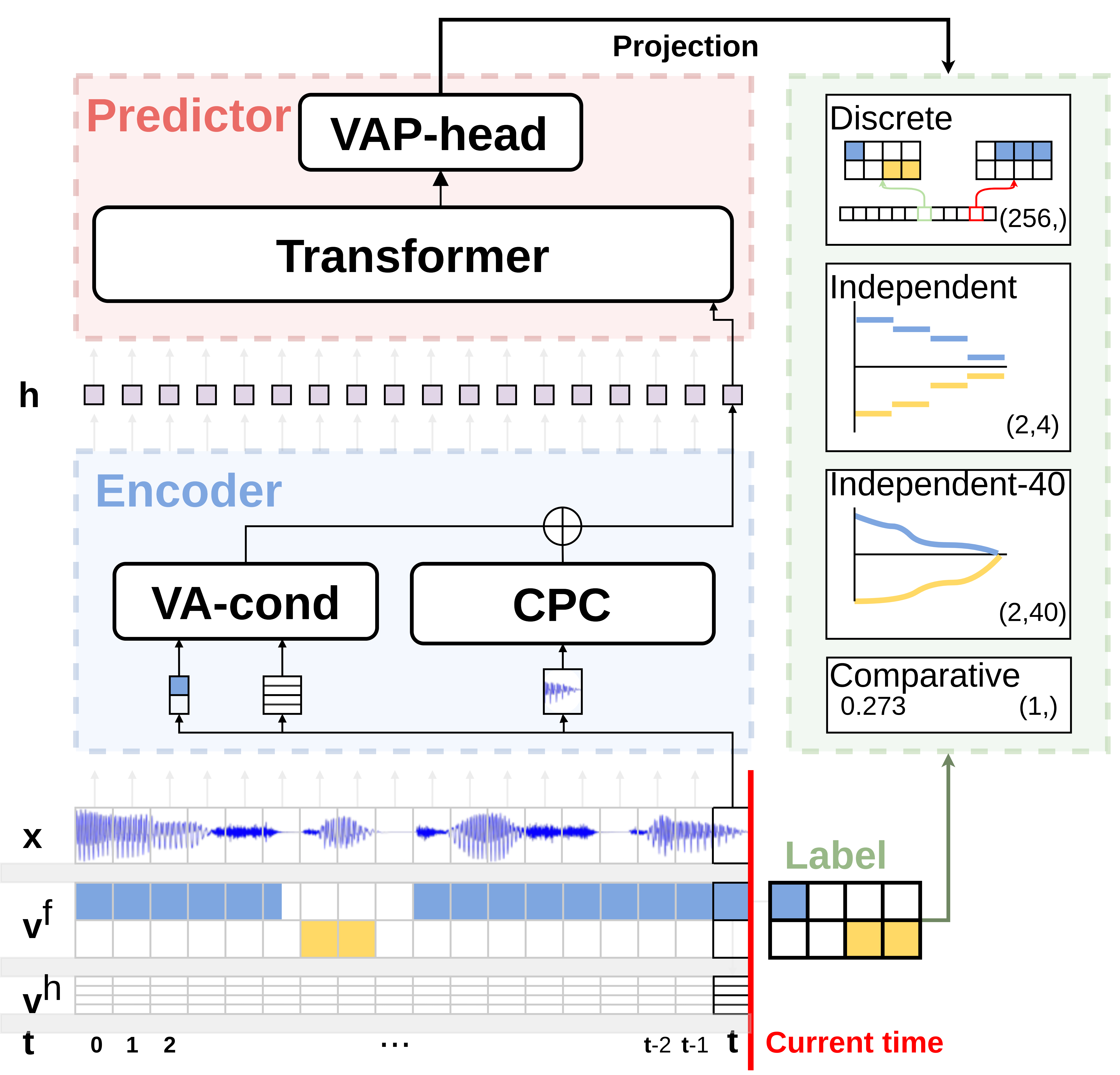}
  \caption{The VAP model processes the input features at time $t$. The input to the model is the combined speech waveforms of the two speakers ($x_t$), the VA-frames of the window ($v^{f}_t$), and the longer VA-history ($v^{h}_t$). The waveform and VA-features are processed separately, projected to a common feature space and added together to produce the predictor input, $h_t$. The predictor consists of a causal transformer feeding into the VAP-head to produce the output projection. The green box illustrates the various outputs of the different models that we compare.}
  \label{fig:model}
\end{figure}

There are two main approaches to data-driven modeling of turn-taking, characterized by the cost of data collection, training objective and \textit{model utility}. We use \textit{model utility} to denote the ease of mapping the model outputs to specific actions, or in other words, how straightforward the model outputs are interpreted as specific actions by the researcher. The training objectives can coarsely be organized in to two groups: \textbf{supervised} and \textbf{self-supervised} learning. Supervised learning refers to classification models which map some input feature space to specific labels. As these labels are typically human-annotated, the cost of data collection is high. On the other hand, since the training objective is very specific, the \textit{model utility} is high, as the labels can be interpreted explicitly. This has been the most commonly used approach for training turn-taking systems in the past \cite{schlangen06, meena, tthuman15}.

Self-Supervised Learning (SSL) can be viewed as the task of reconstructing, or predicting, occluded parts of the data. The idea is to train on these general tasks, minimizing the networks capacity to exploit ``shortcuts", in order to find patterns useful for modeling the underlying data distribution. Once trained, a SSL model can be fine-tuned on a particular dataset, retrained on a downstream task, its representations used as input to other systems, or used directly through ``zero-shot" classification. Zero-shot classification is done by mapping the model's outputs directly to the relevant classes in a specific task, without the need of additional training. Many areas of research have shown promising performance gains by utilizing SSL such as Vision \cite{simclr, vicreg}, Audio \cite{cpc, wav2vec2, hubert} and NLP \cite{gpt3, bert}.  

Compared to supervised learning, data collection for SSL is much cheaper (since no annotation is needed), whereas the general self-supervised training objective produces lower \textit{model utility} outputs. For example, a trained self-supervised model may provide good representations to be used in a downstream task, but the outputs themselves may not be directly interpreted or practically useful (in and of themselves). 

Skantze \cite{vadpred} introduced an incremental and general SSL task suitable for modeling turn-taking, where a system is trained to predict probabilities associated with future VA, for each speaker, over a dialog. The idea is that if a model can successfully predict the future VA of a dialog, that information can be used to control the turn-taking actions of a conversational system. In addition to the training objective, they introduced two zero-shot tasks (i.e., ways of utilizing the low \textit{model utility} outputs), in order to infer turn-taking events directly. In their work they showed that this general task of modeling future VA was successful in producing turn-taking models which could distinguish between SHIFT and HOLD, along with more complex tasks described below. Others reproduced these finding while improving performance by using more advanced input features and model architectures \cite{roddy, roddy2} as well as its generalization across languages \cite{wardpred}.

In this paper, we extend prior work, highlight a theoretical weakness of their design, and propose an alternative. We show empirically that our design improves performance on two new zero-shot tasks, while having equal or greater performance on existing tasks. We provide publicly available code\footnote{\url{https://github.com/ErikEkstedt/vap_turn_taking}} which standardizes the evaluation ``zero-shot" metrics, events (moments of interest), and model output layers, in order to make these tasks more accessible to the research community at large.

\section{Voice Activity Projection}
Projection is the capacity to predict future conversational events as a way for humans, with noticeable slow mechanisms for producing speech, to respond quickly without delay \cite{sacks:74,timing:15,content:15}. 
We thus define Voice Activity Projection (\textbf{VAP}) as the task of predicting the future VA of each interlocutor in a dialog. 

\begin{figure}[ht]
  \centering
  \includegraphics[width=\linewidth]{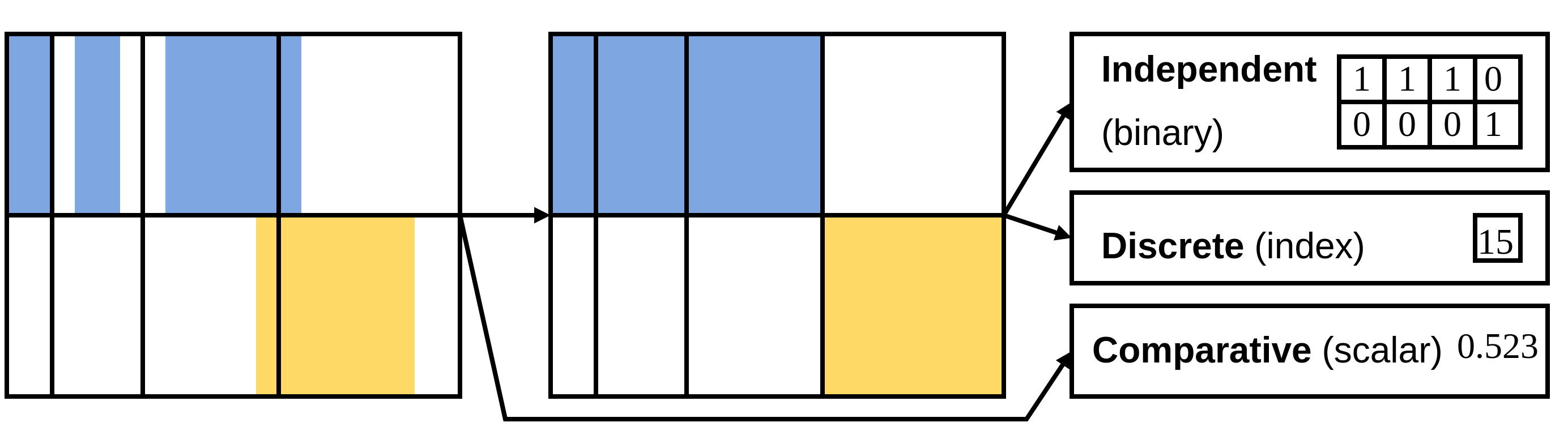}
  \caption{VAP window. Left to right: A window of VA (blue/yellow regions) is organized into 8 bins.
  The VA ratio in each bin is extracted and if it succeeds a specific threshold the bin is considered active. The resulting representation can be modeled directly (Independent) or mapped to a specific state (Discrete). The VA ratio over the entire window, disregarding the bins, can also be modeled directly (Comparative).}
  \label{fig:vap}
\end{figure}

We define a window horizon to construct a VAP window that contains the future VA information to model, at each frame step, over the course of a dialog. The VAP window is discretized into a fixed number of bins, and a VA threshold is used to determine whether the bins are considered active or not, producing the final VAP window used as labels during training, as seen in Figure~\ref{fig:vap}. Previous work \cite{vadpred} used uniformly sized bins of 50ms, covering a total of 3 seconds into the future, and was trained to predict the resulting 60 bins, independently, for each speaker. Once trained, the model would output a value representing the probability of each bin being active or not.

However, given this training objective, it is not clear what the probability associated with each bin means. Considering each bin independently, the probability associated with that bin is of course the probability associated of it being active or not. But if the VAP window as a whole is considered, the interpretation of the outputs becomes less clear. A direct approach could be to take the explicit values of each bin, using their independent probability distributions, and use the resulting projection window as a an explicit VAP state the model deems likely. However, since all bins are independent, this may result in a VAP window state which is heavily out of distribution. Consider a model output where each of the bins contains equal probability of 0.5, does this correspond to all states being equally likely? Are states, highly out-of-distribution, where ``every other" bin is active, as likely as only a single active speaker?

The problem originates from the fact that each bin is modeled independently and cannot be used, in a theoretic way, to statistically model the projection window as whole. Therefore, we propose a change to the task by modeling the entire VAP window directly, keeping the dependence between the VAP window bins. To achieve this, we enumerate each possible configuration of a VAP window as separate states. This way, a VAP window can be viewed as sequence of bits where the total number of states grows exponentially as $2^{n\_bits}$. This exponential growth of possible states restricts the number of bins used to encode a VAP window, and we limit the number of bins to 4 for each speaker, resulting in 8 total bits, or 256 different possible states. The objective for the model is then to output a probability distribution over these states.

Given the practical constraint on the total number of bins, we define a VAP window of 2 seconds, where the bin-sizes increase for bins farther into the future (as we can expect the prediction to be harder, and thus the time resolution lower). Specifically, we use bin durations of 200, 400, 600 and 800ms, as visualized in Figure~\ref{fig:vap}. 

For comparison, we define four different model types, as seen in Figure~\ref{fig:model} and \ref{fig:vap}. The \textbf{Discrete} model is the one we have just described, which outputs a probability distribution over the 256 possible VAP window states. The \textbf{Independent} model uses the same amount of bins, but models the bin activation independently. In order to represent a baseline closer to the original implementation \cite{vadpred}, we also define the \textbf{Independent-40} model, which uses 40 bins of uniform duration (50ms), for each speaker, over the 2 second future time window. As an additional alternative, we define the \textbf{Comparative} model, which simply predicts the ratio of VA of the two speakers directly as a single scalar.

\section{Zero-shot classification of turn-taking events}
We define four tasks that are related to the prediction of turn-taking events,  which we deem relevant for turn-taking in conversational systems \cite{Skantze2021}. As discussed above, in this paper we take a zero-shot approach to these tasks, where we directly map the predictions made by the VAP model to the classes that are to be predicted. Tasks (a) and (d) are adopted from prior work \cite{vadpred}, while (b) and (c) are introduced in this paper. 

\begin{figure}[h]
    \centering
    \includegraphics[width=.48\linewidth]{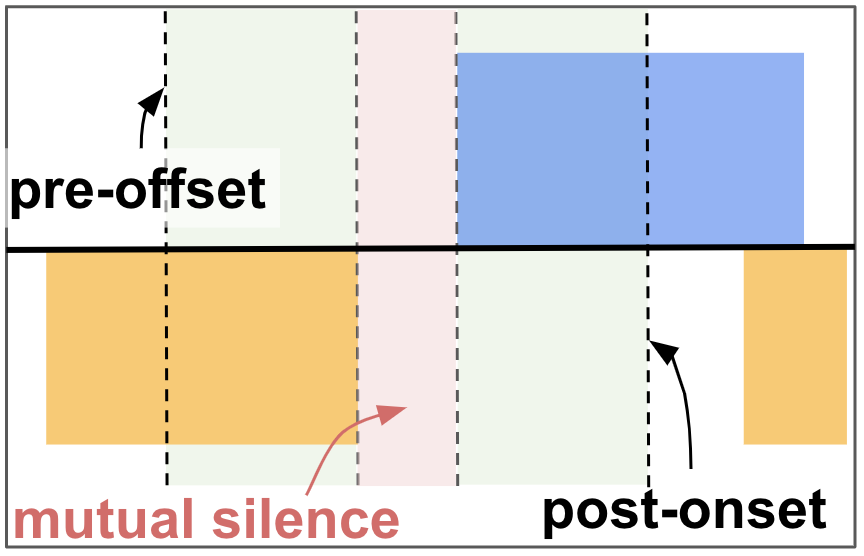}
    \includegraphics[width=.48\linewidth]{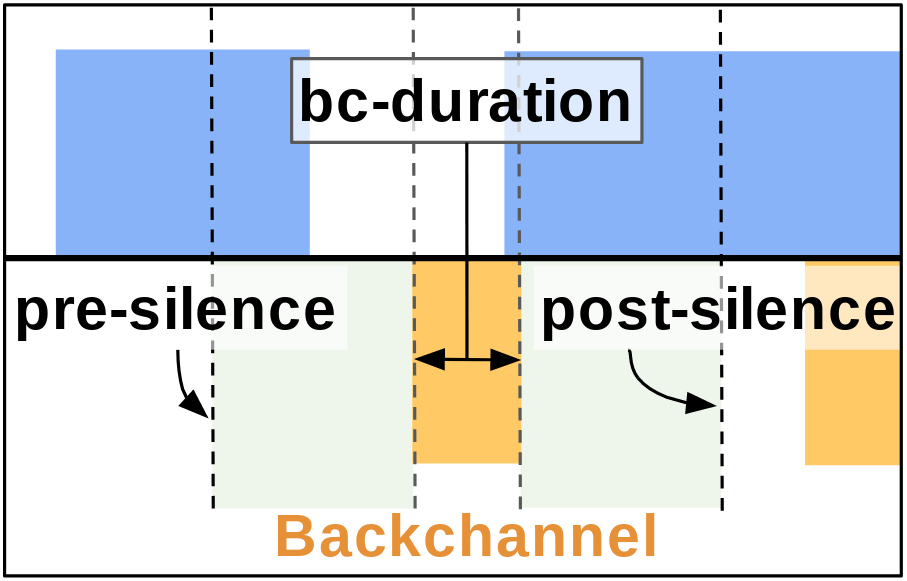} 
    \caption{\textbf{Left}: Only a single speaker can be active in regions (green) before/after a mutual silence (red). SHIFT/HOLD are determined by the previous and next speaker (the example illustrates a SHIFT). \textbf{Right}: The ``backchanneler" cannot be active in the green regions before/after the BC segment. 
    }
    \label{fig:events}
\end{figure}

\textbf{(a) SHIFT vs HOLD (S/H)}: 
This task evaluates how well the model predicts the next speaker during mutual silence, i.e., whether the current speaker will HOLD the turn, or whether the turn will SHIFT to the other speaker. To further limit the analysis to proper turn shifts, we also define two parameters: \textit{pre-offset} (1s) and \textit{post-onset} (1s), which define areas where only a single speaker can be active, as illustrated in Figure~\ref{fig:events} (Left). The frames used for evaluation starts 50ms into the silence, covering a total of 100ms consecutive frames. 

\begin{figure}[h]
    \centering
    \includegraphics[width=\linewidth]{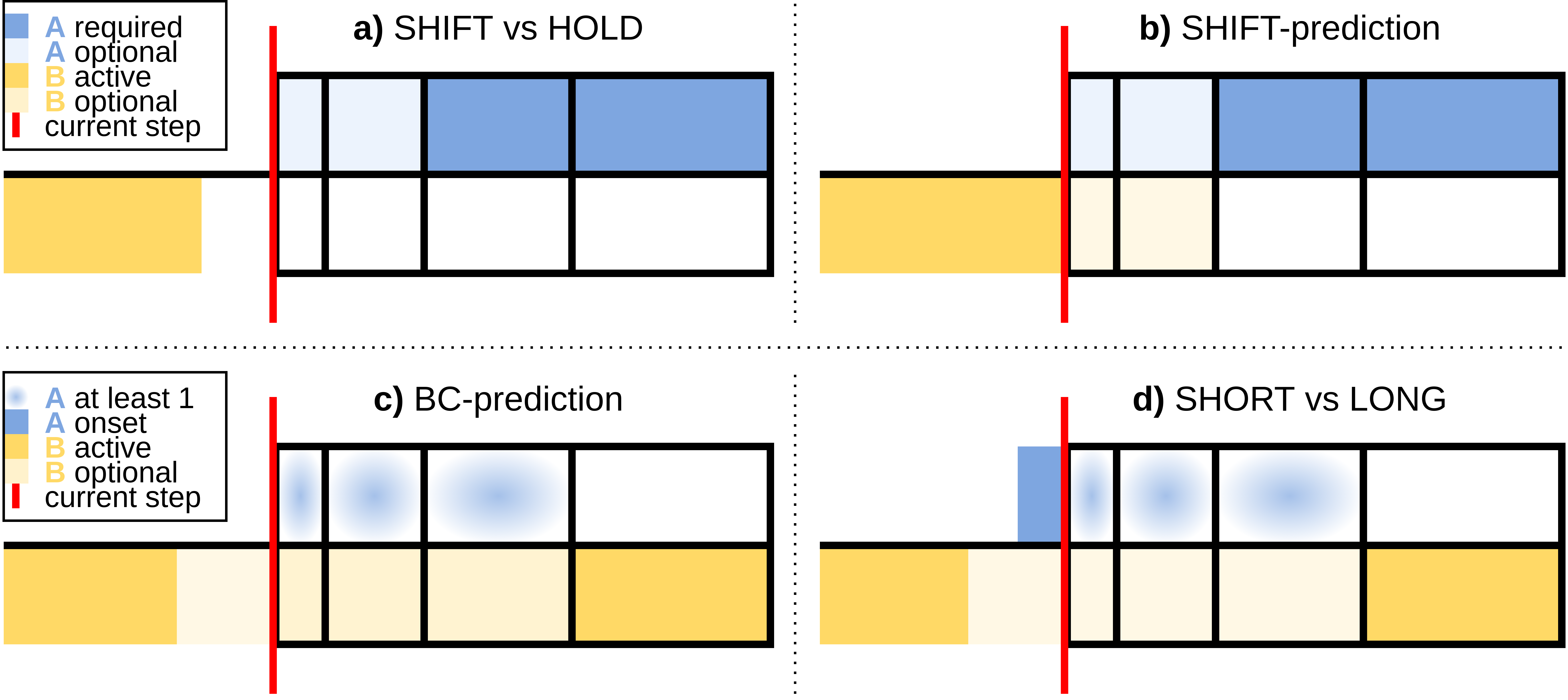}
    \caption{(a) and (b) illustrate the VAP subsets representing A (blue) being the next speaker. The bottom row illustrates the VAP subsets representing a prediction that A (blue) will initiate a BC (c) or initiated a SHORT segment (d).} 
    \label{fig:template}
\end{figure}

To perform zero-shot classification of this task, we mainly consider who the model deems to be the most likely speaker in the immediate future. For the Independent models, we follow prior work \cite{vadpred} and sum together the probability masses over all bins for each speaker, and pick the largest mass to be considered the next speaker. The Comparative model directly outputs a single scalar which is used to find the winner. 
For the Discrete model, we compare two subsets of VAP classes, where only a single speaker is active (with the conditions that the first two bins optionally are active while the last 2 are explicitly active), as illustrated in Figure~\ref{fig:template}a. 

\textbf{(b) SHIFT prediction (S-pred)}:
This task evaluates how well the model can continuously predict an upcoming SHIFT in the near future, while a speaker is still active. Whereas most conversational systems of today only consider the S/H task above, it is well known that humans are also able to predict upcoming turn shifts \cite{timing:15, content:15}, and so we think this is an interesting new task. We consider a range of 500ms that covers the end of a VA segment, before a SHIFT-event (as defined above), as positive samples. Similarly, we sample negative ranges, of the same duration, from regions where a single speaker is active but far away (2s) from future activity of the other speaker.

Zero-shot classification of this task is similar to the S/H task described above. However, for the Independent models, we follow prior work and only consider the bins from the more distant future (omitting the activity in the first 600ms), when doing the comparison. For the discrete model, we include classes which have optional activity of the current speaker, in the near future, to the comparison subsets described above, as illustrated in Figure~\ref{fig:template}b. The \textbf{S-pred} probability for the Comparative model is simply the probability associated with the non-active speaker (the compliment to the active speaker).

\textbf{(c) Backchannel prediction (BC-pred)}: 
This task evaluates how well the model can continuously predict an upcoming BC in the near future (similar to \cite{bc_15, bc_17}). We identify BCs as short and isolated VA segments, as illustrated in Figure~\ref{fig:events} (Right). For a segment to be considered a BC it cannot be longer than \textit{bc-duration} (1s) and there should be no other activity, from that speaker, in a region surrounding the segment, defined by \textit{pre-silence} (1s) and \textit{post-silence} (2s). The final condition is that a BC must be preceded by VA from the other speaker, i.e., a short VA segment from a speaker cannot be considered a BC within the turn of that same speaker. Similarly to the S-pred task, we consider regions of 500ms before a BC as positive samples. The negatives are also sampled similarly to the S-pred task, with the addition of allowing for non-active segments, i.e., backchannels can be predicted during silences as well.

Zero-shot classification for this task is arguable trickier, at least for the Independent models, as we cannot simply compare the future VA probability mass of the two speakers. To project a BC, we expect the non-active speaker to have some short VA in the immediate future, but that the active speaker should have relatively more VA towards the end of the projection window (i.e., continue speaking). Thus, for the Independent models, we require the last bin of the active speaker to be higher than the last bin of the non-active speaker, and that any of the earlier bins for the non-active speaker must be larger than the last bin of the non-active speaker. For the Discrete model, we can simply sum up the probability of the classes that represent this future, as illustrated in Figure~\ref{fig:template}c. It is not clear how the Comparative model could distinguish the prediction of a SHIFT from a BC and is therefore not evaluated on this task.

\textbf{(d) SHORT vs LONG (S/L)}: 
This task considers the onsets of VA segments, after a speaker shift, where the model must distinguish whether the segment is the prefix of a SHORT segment (a BC) or a LONG segment (a proper SHIFT). For a conversational system, this is useful, as it would help to decide whether the system may continue speaking (in case of a BC from the user), or whether it should yield the turn. As positive examples (SHORT), we choose the onset of BCs. As negative examples (LONG), we choose the onset of SHIFTs (as defined above). The evaluation region covers the first 200ms of the onset.

Zero-shot classification for this task is similar to BC-pred, with the exception that who we refer to as the ``non-active" speaker above has just initiated some VA, before the BC projection is made. 

\section{Model \& Training}
In order to investigate whether the proposed VAP objective (i.e., the dependent modeling of the Discrete model) is advantageous over the independence definition of prior work, we define a fixed base model used across experiments. The base model consists of a frame-wise speech and VA encoder followed by a sequence predictor, illustrated in Figure~\ref{fig:model}. 

The encoder consists of two sub-modules where a speech module processes raw waveforms, $x$, directly using a pre-trained CPC \cite{cpc_across} model that outputs frame representations $h_{speech,t}\in\mathbb{R}^{256}$, at 100Hz. A second VA module, matching the frame rate of the speech encoder, processes the current VA frame vector $v^f_t\in \{0, 1\}^2$, along with a concise representation over the VA history. The VA-history features provides long ranging contextual information outside of the receptive field of the acoustic model. This history is defined as the activity ratio of speaker A over speaker B for regions of size \{-inf:60, 60:30, 30:10, 10:5, 5:0\} seconds in to the past, where 0 is the current time step, resulting in a vector $v^h_t\in\mathbb{R}^5$ with values between 0 and 1, for each frame. The VA module projects the VA features to vectors $h_{va,t}, h_{his,t} \in \mathbb{R}^{256}$ which are added to the speech representation $h_{speech,t}$ to produce the encoder output $h_t$, for each frame $t$. The predictor consists of a causal, decoder only, transformer \cite{transformer} using a hidden size of 256, 4 layers, 4 heads and 0.1 dropout. The output of the transformer model is fed to the VAP-head, a final linear layer, which outputs logits associated with the respective model types. The models differs by the size of the logits which are $(256,), (2, 4), (2, 40)$ and $(1,)$ for the Discrete, Independent, Independent-40 and Comparative models respectively. The discrete model is optimized through the cross-entropy loss, while the others use the binary-cross-entropy loss. 

We train on the Switchboard dataset \cite{swb} containing a total of 2438 dialogs. We exclude 98 dialogs which contain more than 2 speakers
The remaining dialogs are split into a test-set of 135 dialogs (used in the final evaluation) and a training-set of 2205 dialogs. The training-set is organized into k-fold splits to create 11 pairs of train/validation sets of 2000/205 dialogs. The dialog waveforms are volume normalized, resampled to 16kHz and, unlike prior work \cite{vadpred}, mixed to a single channel. The audio waveforms are split into 10s chunks (with 2s overlap) along with the corresponding VA-features to produce the input to the models. We train with an early stopping criteria of 20 epochs, on the weighted F1 score of the S/H measure. We use the AdamW \cite{adam, opt_decoupled} optimizer with a learning rate of 3.63e-4, found using pytorch-lightning's learning rate finder \cite{lr_finder}. The models and training were implemented in python using the PyTorch \cite{pytorch} and PyTorch-Lightning \cite{pl} libraries and are available online\footnote{\url{https://github.com/ErikEkstedt/conv_ssl}}.

\section{Results}
We extract the average loss for the best Discrete and Independent models over the validation set (k-fold 0) to investigate how it varies depending on the number of frames available as context. We note a higher aggregate loss at the start of the samples (since less context is available) and choose a threshold of 3s, where the loss decreases to values approximately consistent over the remaining frames, as a point of minimum context for the evaluation events. We iterate over the validation set once, for each trained model, to find threshold values for the S/L, S-pred and BC-pred tasks e.g. a task probability exceeding the threshold yields a positive prediction. Given the appropriate thresholds we evaluate each model over the test set to extract the weighted F1-score for each task. The final scores, displayed in Table~\ref{tab:result}, are the average performance for each model type. 

\begin{table}[h]
    \caption{Turn-taking metrics. The average weighted F1-score over 11-kfold splits for the different models. (*) indicates statistically significant improvement over alternatives. The \textbf{(S)} in the \textbf{S/H} column represent the F1-score for the \textbf{SHIFT} label.} 
    \label{tab:result}
    \centering
    \begin{tabular}{l|l|l|l|l}
        \toprule
        \textbf{Model}&\textbf{S/H (S)}&\textbf{S/L}&\textbf{S-pred}&\textbf{BC-pred}\\
        \hline
        Discrete (ours)& \textbf{.899 (.510)} & \textbf{.786} & \textbf{.733}* & \textbf{.723}*\\
        Independent& .897 (.486) & \textbf{.786} & .718 & .685\\
        Independent 40& .896 (.483) & .778 & .712 & .661\\
        Comparative& .893 (.475) & .546 & .714 & N/A\\
        Majority class& .843 (0) & .565 & .333 & .333\\
        \bottomrule
    \end{tabular}
\end{table}

We perform a 2-sided ANOVA test (using scipy \cite{scipy}) to find if any model's performance is statistically significantly different from the others. After the ANOVA test, we conduct an ad-hoc t-test, comparing our proposed model (the best scoring across tasks) to the alternatives. We note that the performance on the S/H and S/L tasks show no significant difference. However, on the prediction tasks we get a statistically significant ($p < 0.025$) difference where the proposed model outperforms the alternatives, denoted by ``*" in Table~\ref{tab:result}.

It is interesting to note that the task that benefits most from the Discrete model is the BC-pred task, which is also the task that has the most complex dependency assumptions of the speakers' future VA. 

\section{Conclusion and Future Work}
The results show that our proposed objective is favorable for modeling turn-taking when considering more complex predictive tasks. The proposed objective show equal or greater performance on zero-shot tasks defined in prior work (\textbf{SHIFT/HOLD, SHORT/LONG}) while being statistically significantly better on two new predictive tasks (\textbf{SHIFT-prediction, Backchannel-prediction}). 

We argue that SSL and general training objectives is a promising direction for modeling conversational dynamics such as turn-taking. In this work we highlight a theoretical shortcoming of earlier work and propose a more theoretically sound alternative. Instead of modeling the future VA of the two speakers independently, we propose an objective which considers an entire VAP window as a whole. On a qualitative note, the output of our proposed design increases \textit{model utility} by providing well defined outputs that correspond to specific future states. 

The results indicate that future work, related to the modeling of more complex conversational dynamics, can benefit from our proposed VAP objective. For future work, we plan to investigate additional projection window representations, using varying forms on information (e.g., prosodic), to model the dynamics of conversation, in addition to the voice activity projection task explored in this paper. Videos and extra resources are publicly available\footnote{\url{https://erikekstedt.github.io/VAP/}}.

\clearpage

\bibliographystyle{IEEEtran}
\bibliography{references}
\end{document}